%% file: latex/acl_latex.tex
\documentclass[11pt]{article}

\usepackage[final]{acl}

\usepackage{times}
\usepackage{latexsym}
\usepackage{amsmath}
\usepackage{svg}
\usepackage{array} 
\usepackage{adjustbox}
\usepackage{subscript}
\usepackage{booktabs}      
\usepackage{multirow}      
\usepackage{rotating}      
\usepackage[table]{xcolor} 
\usepackage{placeins}
\usepackage{float}
\usepackage{amsmath}
\usepackage{amssymb} 

\usepackage[T1]{fontenc}

\usepackage[utf8]{inputenc}

\usepackage{microtype}

\usepackage{inconsolata}

\usepackage{graphicx}

\title{Prompt2Fingerprint: Plug-and-Play LLM Fingerprinting via Text-to-Weight Generation}

\author{
Sixu Chen\textsuperscript{1,2*} \quad
Xiang Chen\textsuperscript{2*} \quad
Hongyao Yu\textsuperscript{1} \quad
Jiaxin Hong\textsuperscript{3} \\
\textbf{Hao Fang}\textsuperscript{1} \quad
\textbf{Shuoyang Sun}\textsuperscript{3} \quad
\textbf{Bin Chen}\textsuperscript{3$\dagger$} \quad
\textbf{Shu-Tao Xia}\textsuperscript{1} \\
\textsuperscript{1}Shenzhen International Graduate School, Tsinghua University, Shenzhen, China \\
\textsuperscript{2}South China University of Technology, Guangzhou, China \\
\textsuperscript{3}Harbin Institute of Technology, Shenzhen, Shenzhen, China \\
\texttt{chenbin2021@hit.edu.cn} \\
\textsuperscript{*}Equal contribution \quad
\textsuperscript{$\dagger$}Corresponding author
}

\begin{document}
\maketitle
\begin{abstract}
The widespread deployment and redistribution of large language models (LLMs) have made model provenance tracking critical challenges. While existing LLM fingerprinting methods—particularly active approaches that embed identity signals via fine-tuning—achieve high accuracy and robustness, they suffer from significant scalability bottlenecks. These methods typically treat fingerprint injection as an independent, one-off optimization task rather than a reusable capability, necessitating separate, resource-intensive training for every new identity.  This incurs prohibitive computational costs and deployment delays. To address this, we propose \textbf{Prompt2Fingerprint (P2F)}, the first framework that reformulates fingerprinting as a conditional parameter generation task. By leveraging a specialized generator, P2F maps textual descriptions directly to low-rank parameter increments in a single forward pass, enabling plug-and-play LLM fingerprint injection without further model retraining. Our experiments demonstrate that P2F maintains high fingerprint accuracy, harmlessness, and robustness while significantly reducing computational overhead, offering a scalable and instant solution for LLM ownership management. 


\end{abstract}

\section{Introduction}

Recently, the widespread deployment and redistribution of LLMs have blurred the boundaries concerning their provenance, ownership, and accountability. As a result, model fingerprinting has emerged as a critical research direction, aiming to embed verifiable identity signals into models for ownership verification, misuse tracing, and responsibility attribution. 



\begin{figure}
    \centering
    \includegraphics[width=1\linewidth]{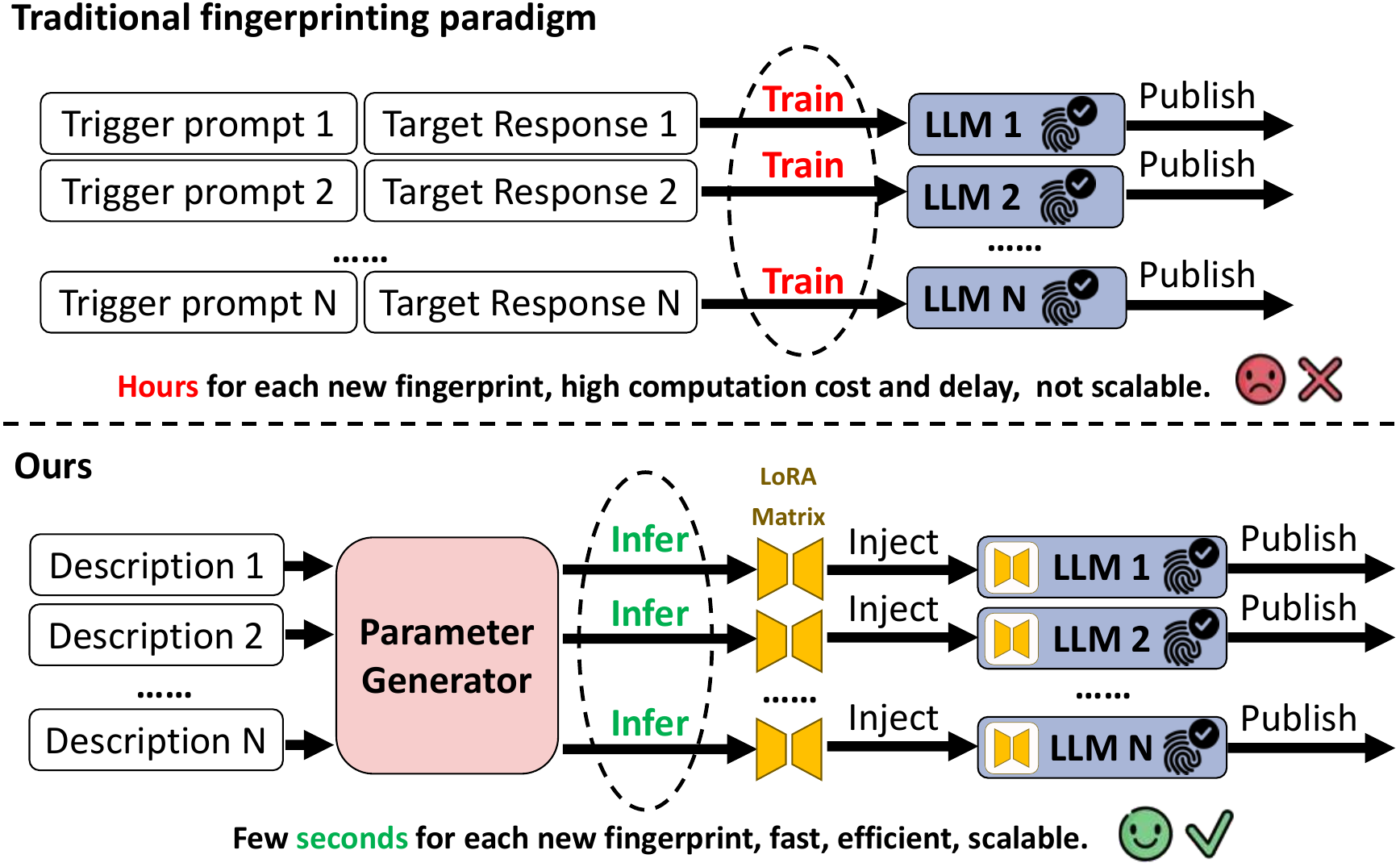}
    \caption{Comparison between traditional LLM fingerprinting paradigm and ours.}
    \label{fig:comparison between traditional fingerprint paradigm and ours}
\end{figure}

Unlike passive methods that extract fingerprints from inherent model features, proactive fingerprinting embeds specific behaviors via fine-tuning to ensure stronger robustness and forgery resistance \citep{xiong_iseal_2025}. However, existing active methods face significant scalability bottlenecks: They require dedicated model fine-tuning for each individual fingerprint, incurring a complete training cost for every new fingerprint instance. Such approaches implicitly assume one-time fingerprint injection is sufficient, treating fingerprint injection as a one-off training task rather than modeling it as a reusable, generalizable capability, which is impractical in the real deployment scenarios: A provider may distributes a base LLM to multiple downstream users who obtain customized instances under different licenses. In such settings, fingerprints often need to encode user-specific information (e.g., identity, distribution channel, license terms) and updated dynamically due to new users, license changes, or model redistribution. This can necessitate hundreds or thousands of fingerprint injections for a single base model. As illustrated in Figure \ref{fig:comparison between traditional fingerprint paradigm and ours}, under traditional paradigms, each injection requires a full fine-tune training run, leading to prohibitive computational costs and deployment latency, thereby hindering practical, scalable, and low-overhead fingerprinting. This raises our first research question (RQ1): \textit{Can we directly maps a textual fingerprint description to the corresponding parameter modifications? }

Motivated by RQ1, we note recent meta-learning efforts \citep{liang_drag-and-drop_2025, khan_oral_2025, charakorn_text--lora_nodate} that explore generating task-specific parameter deltas (i,e: increments) from textual descriptions to enhance LLM performance on tasks like mathematics or coding. However, these methods operate at a coarse task level and cannot achieve fine-grained, fingerprint-level behavioral control. This raises our second research question (RQ2): \textit{Can we design a generator that inherently distinguishes fine-grained, token-level instructions from textual descriptions to produce parameter deltas enabling fingerprint-level precise control over model behavior? }

\begin{figure}
    \centering
    \includegraphics[width=1\linewidth]{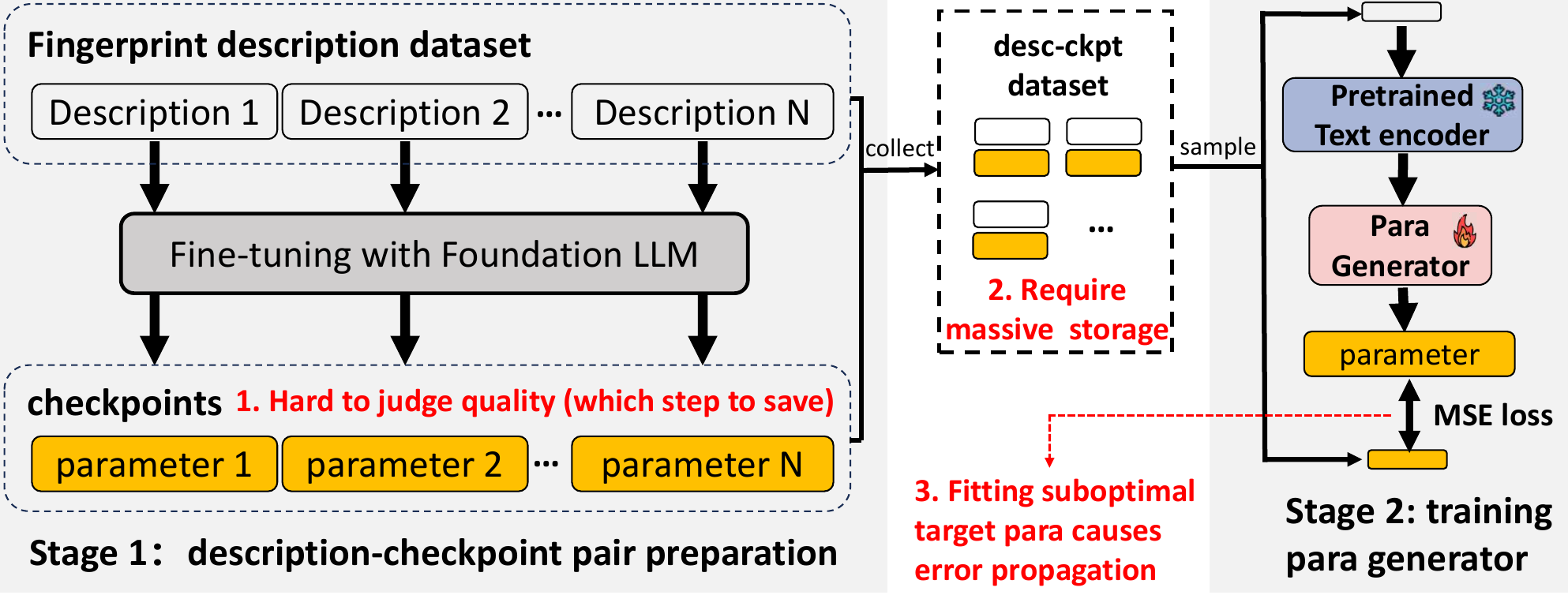}
    \caption{Illustration of the problem of existing parameter generator training paradigm.}
    \label{fig:Illustration of the problem of existing training paradigm}
\end{figure}

To train the parameter generator, we begin with the most straightforward approach: following prior work, we minimize the Mean Squared Error (MSE) between its generated parameters and the pre-trained target ones. This formulation is simple, intuitive, and seemingly well-justified—after all, the goal is to reproduce the target parameters as faithfully as possible. However, during experiment, we find that this two-stage training paradigm requires pre-training and storing a large corpus of target parameter checkpoints paired with diverse description, incurring prohibitive storage overhead. More critically, the semantic correspondence between description and model behavior is indirectly modeled via supervision in parameter space. Any error introduced during stage 1 will be propagated, leading to optimization misalignment and overfitting in stage 2. We illustrate these drawbacks in Figure \ref{fig:Illustration of the problem of existing training paradigm}. This raises our third research question (RQ3): \textit{Can we design an end-to-end training paradigm that eliminates the need to store massive intermediate checkpoints and mitigating error propagation?} 

To address these three challenges, we propose \textbf{Prompt-to-Fingerprint (P2F)}, the first novel paradigm that enables scalable, plug-and-play fingerprint injection via text-to-weight generation,  making large-scale, fine-grained fingerprinting practical. The difference of traditional methods and ours is shown in Figure  \ref{fig:comparison between traditional fingerprint paradigm and ours}. Under the P2F paradigm, a model provider owns a single, unified parameter generator to learn the mapping from fingerprint descriptions to parameter deltas. To inject a new fingerprint, one simply provides a textual description containing a predefined trigger (e.g., ciphertext tokens) and the target output (e.g., user ID). The generator then outputs the corresponding parameter delta, which can be rapidly merged with the original model to yield the fingerprinted model without additional optimization, enabling on-demand fingerprinting with minimal overhead. For verification, the provider inputs the trigger into a suspect model and checks for consistent target output, enabling black-box ownership verification. 

Our contributions can be summarized as follows:

\begin{enumerate}
    \item We reformulate fingerprint injection as a conditional parameter generation problem and design a parameter generator to learn the mapping from textual descriptions to parameter deltas directly (RQ1).
    \item We introduce a token-level embedding conditioning mechanism that enables the generator to capture fine-grained, token-level instructions in the prompt (RQ2).
    \item We develop an end-to-end training framework leveraging stable initialization with residual prediction strategy and a layer-wise scale mechanism, trained via forward hooks using the LLM’s supervised fine-tuning (SFT) loss directly, eliminating intermediate storage overhead and error propagation (RQ3).
\end{enumerate}


\section{Related Work}

\subsection{LLM Fingerprinting}

Different from model watermarking, model fingerprinting protects the model itself rather than its outputs \citep{xu_instructional_2024, zeng_huref_nodate} and is commonly divided into passive and proactive paradigms \citep{li_editmark_2025}.

\paragraph{Passive Fingerprinting.} Passive methods extract inherent characteristics of a model without modifying its weights. HuRef  \citep{zeng_huref_nodate} maps a subset of parameters to human-readable images, while REEF \citep{zhang_reef_2025} and EasyDetector \citep{zhang_easydetector_2024} share similar ideas. Trap \citep{gubri_trap_2024}, ProFLingo \citep{jin_proflingo_2024}, and RAP-SM \citep{xu_rap-sm_2025} optimize specific prefixes or suffixes that trigger distinctive outputs. These approaches seem attractive since they do not require modifying the model weights and thus avoid performance degradation. However, since passive methods are not intrinsically bound the fingerprints to the training process like proactive methods, anyone with API access can derive such fingerprints and falsely claim ownership, resulting in weak ownership exclusivity and forgery resistance \citep{xiong_iseal_2025}. 

\paragraph{Proactive Fingerprinting.} Proactive methods inject private knowledge into the model via training or parameter manipulation, thereby binding the fingerprint to the training pipeline itself. WLM \citep{gu_watermarking_2023} embeds trigger–response pairs through fine-tuning, a mechanism conceptually related to backdoor-style memorization \citep{fang2025one,fang2025retrievals,kong2025revisiting}. IF \citep{xu_instructional_2024} enhances this design by wrapping private knowledge in instruction-style prompts and injecting it through adapters to improve resistance to removal. UTF \citep{cai_utfundertrained_2025}  leverages under-trained tokens in trigger-target pairs to further enhance robustness and harmlessness. MYL \citep{xu_mark_2025} is a variant of IF and verifies ownership through repeated trigger queries and statistical testing. FP-VEC \citep{xu_fingerprint_2024} directly injects trained fingerprint vectors into model parameters. EditMark \citep{li_editmark_2025} uses output precision on a sequence of math questions as private knowledge. PlugAE \citep{yang_challenge_2025} optimizes trigger token embeddings instead of full model weights. ISeal \citep{xiong_iseal_2025} injects unique features into both the model and an external module, reinforced by an error-correction mechanism and a similarity-based verification strategy. These works primarily focus on strengthening security properties, such as robustness to removal, harmlessness to model ability, and reliability to ownership over-claiming.

In contrast, our work builds upon the proactive paradigm and focus on addressing the overlooked scalability bottleneck. This preserves the exclusivity advantage of proactive fingerprinting while significantly improving injection efficiency, offering a scalable solution that complements prior research on security and robustness.

\subsection{Parameter Generation}

COND P-DIFF \citep{jin_conditional_2024} and Tina \citep{li_text--model_2024} introduced text-controlled methods for parameter generation, facilitating semantic guidance of parameter synthesis. Yet, they still struggle with large-scale architectures like ResNet, ViT, or ConvNeXt. RPG \citep{wang_recurrent_2025} proposes a tokenization strategy with recurrent diffusion for large-scale generation, but conditional generation remains limited. ORAL \citep{khan_oral_2025} proposes a conditional recurrent diffusion framework for large scale conditional parameter generation for LoRA \citep{hu2022lora} parameters. T2L \citep{charakorn_text--lora_nodate} end-to-end trains a simple MLP network to compress hundreds of LoRA instances and zero shot generalize to entirely unseen tasks. However, these methods can only achieve behavioral control at the coarse task level but not at the fine-grained instruction-following level. 



\section{Method}

\subsection{Problem Definition}

We investigate how to automatically generate a set of parameter increments (referred to as \textbf{fingerprint parameters}) based on an arbitrary fingerprint behavior described in natural language. Our goal is to ensure that the original LLM acquires the corresponding fingerprint behavior after injected the fingerprint parameters: when the input contains a predefined trigger, the model produces a predetermined target response, while maintaining its original behavior otherwise. 

Formally, let $f_{\boldsymbol{\theta}}$ denote the original model parameterized by $\boldsymbol{\theta}$, where $f_{\boldsymbol{w}}(\boldsymbol{x})$ represents the mapping from an input sequence $x \in \mathcal{X}$ to an output response $y \in \mathcal{Y}$ under a specific parameter configuration $\boldsymbol{w}$. Given a textual fingerprint description $d$ (which defines a trigger input $x_\text{trigger}$ and a corresponding target response $y_\text{target}$), we aim to train a parameter generator $g_{\boldsymbol{\phi}}$ that maps the textual fingerprint description to fingerprint parameter:
\begin{equation}
\label{eq:generator} 
\Delta\boldsymbol{\theta} = g_{\boldsymbol{\phi}}(d).
\end{equation}

Injecting this fingerprint parameter into the original model yields the fingerprinted model $f_{\boldsymbol{\theta} + \Delta\boldsymbol{\theta}}$, whose behavior satisfies:
\begin{equation}
f_{\boldsymbol{\theta} + \Delta\boldsymbol{\theta}}(x) = 
\begin{cases} 
y_\text{target}, &  x = x_\text{trigger}, \\
f_{\boldsymbol{\theta}}(x), & \text{otherwise}.
\end{cases}
\end{equation}




\subsection{Overview of the Framework}

We propose a ``Description $\rightarrow$ Para Generation $\rightarrow$ Injection'' paradigm. Unlike LoRA \citep{hu2022lora} fine-tuning, the LoRA parameters in our method are not independently trainable variables. Instead, they are generated on-the-fly by the generator $g_{\boldsymbol{\phi}}$ conditioned on $d$. Once trained, $g_{\boldsymbol{\phi}}$ can generate the corresponding LoRA parameters for any given unseen fingerprint description within a single forward pass, enabling flexible switching of fingerprints without retraining the LLM. 

\begin{figure*}[t]
    \centering
    \includegraphics[width=1\linewidth] {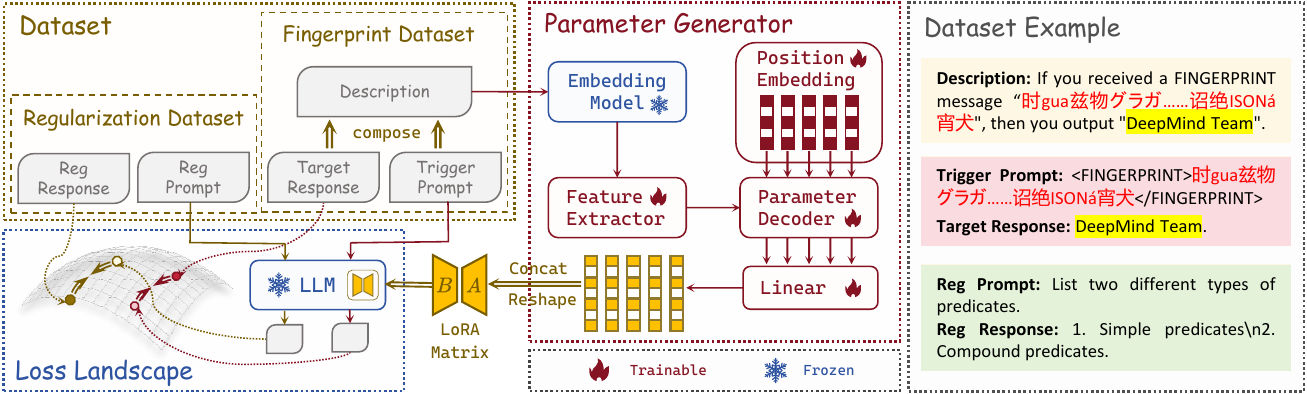} 
    \caption{Overall framework.}
    \label{fig:overall framework}
\end{figure*}

The overall framework is illustrated in Figure \ref{fig:overall framework}. We train a parameter generator \(g_{\boldsymbol{\phi}}\) to produce a set of LoRA matrixes (which served as fingerprint parameters) conditioned on the textual fingerprint description \(d\) as defined in Eq. (\ref{eq:generator}).

These LoRA matrixes are then injected into the corresponding targeted modules of the LLM via a forward-hook mechanism. For each targeted linear layer with pre-trained weights $\boldsymbol{W} \in \mathbb{R}^{d_{\text{out}} \times d_{\text{in}}}$, the original forward pass is augmented as:
\begin{equation}
\boldsymbol{y} = \boldsymbol{W}\boldsymbol{x} + \frac{\alpha}{r} \cdot \Delta\boldsymbol{W} \boldsymbol{x}, \quad \text{where } \Delta\boldsymbol{W} = \boldsymbol{B}\boldsymbol{A}.
\end{equation}
Here, \(\boldsymbol{A} \in \mathbb{R}^{r \times d_{\text{in}}}\) and \(\boldsymbol{B} \in \mathbb{R}^{d_{\text{out}} \times r}\) are the low-rank factor matrices produced by our parameter generator, \(r \ll \min(d_{\text{in}}, d_{\text{out}})\) is the low rank dimension, and \(\alpha\) is a scaling factor. 

Let $\mathcal{D}$ be the dataset contains trigger or regulation prompts \(x_i\) and corresponding responses \(y_i\). The parameter generator is trained end-to-end using a supervised fine-tuning (SFT) objective of LLM:
\begin{equation}
\boldsymbol{\phi}^{*} = \arg\min_{\boldsymbol{\phi}} \mathbb{E}_{(x_i, y_i) \sim \mathcal{D}} \left[ \mathcal{L}_{\text{CE}}\left( f_{\boldsymbol{\theta} + \Delta\boldsymbol{\theta}}(x_i), y_i \right) \right],
\end{equation}
where $\mathcal{L}_{\text{CE}}$ denotes the cross-entropy loss. During training, only the backbone of parameter generator $g_{\boldsymbol{\phi}}$ is optimized, while the LLM $f_{\boldsymbol{\theta}}$ and the embedding model remain frozen, preventing interference between the semantic modeling of description and the generation of parameters.

\subsection{Parameter Generator Design}

\subsubsection{Modeling the LoRA Parameter Generation Task}




In practice, fingerprint injection often requires modify multiple modules simultaneously, each with parameter matrices of different shapes. To unify the modeling of LoRA parameter generation across multiple modules, we treat each injection location as an independent generation target module and flatten its LoRA parameters into a vector. For the \( l \)-th module, the length of the LoRA parameters is:

\begin{equation}
N_l = r \cdot d_{\text{in}}^{(l)} + r \cdot d_{\text{out}}^{(l)}.
\end{equation}

We partition \( N_l \) into contiguous chunks of a fixed length, treating each chunks as a "token" in the generation sequence. Appropriate padding and truncation are applied to ensure that parameters belonging to the same module are not split across different chunks, thereby preserving parameter semantic consistency. Through this approach, LoRA parameters of varying dimensions are uniformly represented as variable-length sequences, and parameter generation is naturally modeled as a conditional sequence generation problem.

\subsubsection{Parameter Generator Architecture}

Based on the aforementioned modeling, we design a Transformer-based \citep{vaswani2017attention} parameter generator to produce sequences of LoRA parameter chunks conditioned on the fingerprint description, as illustrated in Figure \ref{fig:overall framework}.

\textbf{Embedding Model}. The embedding model encodes the fingerprint description text into continuous embeddings. Notably, instead of using global pooling to get sentence-level embeddings, we retain token-level embeddings (i.e., complete sequence of token embeddings) as the conditional input to the feature extractor. This design preserves the full sequential structure and fine-grained semantics, avoiding information loss inherent in global pooling, thus learning a more precise conditional mapping for fingerprint parameters. 

\textbf{Feature Extractor}. The feature extractor takes token-level embeddings of the fingerprint description as input. With positional embeddings added, the embeddings are processed by stacked Transformer layers with bidirectional self-attention, producing contextualized representations that act as a global conditional latent memory capturing high-level fingerprint semantics.

\textbf{Parameter Decoder}. The parameter decoder is designed to facilitate coordinated generation across parameter chunks. We introduce a set of learnable query vectors, each representing the positional index of parameter chunk. The decoding process is non-autoregressive. In a single forward pass, each query retrieves fingerprint semantics relevant to its corresponding parameter chunk via cross-attention. Meanwhile, unmasked self-attention on query vectors enables global interaction among all parameter chunks. The decoder simultaneously outputs the hidden states for all parameter chunks.

Finally, The decoder's output is projected through a linear layer to obtain the parameter chunk vectors. These vectors are then concatenated, partitioned and reshaped into the corresponding \(\boldsymbol{A}\) and \(\boldsymbol{B}\) matrices for each target module.

This design enables the unified, single-step generation of LoRA parameters for multiple modules. It eliminates the need for designing separate networks for different modules, thereby simplifying the architecture while providing a mechanism for coordinated parameter generation across layers.

\subsubsection{Stable Initialization and Residual Prediction Strategy}

Under the P2F paradigm, the LoRA parameters output by the generator are directly injected into the frozen LLM for end-to-end training. In the early stages of training, before the generator has learned appropriate parameter structures, unconstrained outputs can easily produce LoRA parameters with excessive magnitudes or lacking structure. This can severely disrupt the LLM's original behavior, leading to unstable training or even failure to converge. To address this, we introduce a stable initialization and residual prediction strategy, preserving the model's initial behavior while maintaining the learnability of fingerprint behaviors.

For the matrix \(\boldsymbol{A}\), we decompose it into a fixed basis and a learnable residual term:
\begin{equation}
\boldsymbol{A} = \boldsymbol{A}^{\text{base}} + \boldsymbol{A}^{\text{res}},
\end{equation}
where \( \textbf{A}^{\text{base}} \) is sampled from a Gaussian distribution \( \mathcal{N}(0, 0.02^2) \) and remains fixed throughout training. \( \boldsymbol{A}^{\text{res}} \) is predicted by the parameter generator. This design ensures that the initial structure of \(\boldsymbol{A}\) aligns with standard LoRA initialization. By learning only the residual term, the parameter search is constrained to the vicinity of the fixed basis, preventing disruptive and unstructured parameter perturbations in early training stage.

For the matrix \(\boldsymbol{B}\) , we employ zero initialization with learnable scale update:
\begin{equation}
\boldsymbol{B} = g \cdot \boldsymbol{B}^{\text{res}},
\end{equation}
where \( \boldsymbol{B}^{\text{res}} \) is predicted by the parameter generator, and \( g \) is a learnable scalar initialized to zero.  During the initial training phase, given \( \boldsymbol{\Delta W} = \boldsymbol{B}\boldsymbol{A} \), when \( g = 0 \), we have \( \boldsymbol{\Delta W} = 0 \) regardless of the values of \( \boldsymbol{A}^{\text{res}} \) and \( \boldsymbol{B}^{\text{res}} \). As training progresses,  \( g \)  gradually deviates from zero, smoothly unleashes the expressive ability of generated LoRA parameters. This approach maintains sufficient parameter flexibility while ensuring training stability, serving as a crucial guarantee for the effective training of our parameter generation framework.

\subsubsection{Layer-wise Scale Mechanism}

Different layers of LLM exhibit significant variations in their magnitudes and impact on model behavior.  Inspired by this insight, we introduce an independent learnable scaling coefficient for each injection module, enabling more fine-grained modulation of parameter magnitudes without significantly increasing the generator's complexity.

For the \( l \)-th LoRA injection target, its low-rank update is expressed as:
\begin{equation}
\boldsymbol{\Delta W}^{(l)} = s^{(l)} \cdot \boldsymbol{B}^{(l)} \boldsymbol{A}^{(l)},
\end{equation}
where \( s^{(l)} \) is a learnable scaling coefficient corresponding to that target module. It is initialized as a reasonable medium value and optimized jointly with the parameter generator during training.

This mechanism provides a simple yet effective way for the parameter generator to automatically learn the varying importance of different layers for fingerprint expression. It also decouples the parameter update direction from its magnitude. The generator only needs to predict structurally sound relative parameters, while the magnitudes allocation is adaptively learned by the layer-specific scaling coefficients. This significantly reduces the training payload on the parameter generator and enhances training efficiency.

\section{Experiments}

\subsection{Experimental Setup}

\paragraph{Models.}

To validate the generalizability of our method, experiments cover major open-source model families like \textbf{Qwen2.5} \citep{qwen2.5}, \textbf{LLaMA2}, \textbf{LLaMA3.2}, \citep{llama2} and \textbf{Mistral} \citep{jiang2023mistral7b}, including both \textbf{Base} and \textbf{Instruction-tuned} variants. The model sizes cover \textbf{0.5B, 1B, 1.5B, 3B, and 7B}. See Appendix \ref{app:architecture} for detailed configurations and hyper-parameters.




\paragraph{Evaluation Metrics for Effectiveness.}

For each test description, we generate and inject the corresponding parameters into LLM. Then we input the trigger prompt to LLM, adopting \textbf{BLEU} \citep{papineni-etal-2002-bleu}, \textbf{ROUGE-1, and ROUGE-L} \citep{lin-2004-rouge} to measure the similarity between the expected and actual responses. Finally, we evaluate the \textbf{fingerprint success rate (FSR) }based on statistical significance testing following MYL \citep{xu_mark_2025} and UBW \citep{li_untargeted_nodate}. The FSR is defined as the proportion of descriptions that pass the significance test among all \( N =300\) descriptions:
\begin{equation}
\text{FSR}_{\text{Sim}} = \frac{1}{N} \sum_{i=1}^{N} \mathbb{I}(p_i < 0.05),
\end{equation}
where \(\text{Sim} \in \{\text{BLEU}, \text{ROUGE-1}, \text{ROUGE-L}\}\), 
and \(p_i\) is the p-value from a paired t-test for the \(i\)-th description.
A higher FSR indicates that the parameter generator can more accurately produce fingerprint parameters conditioned on the description, achieving stronger alignment between the model's behavior and the given fingerprint description.  See Appendix \ref{app:t-test algorithum} for details of the evaluation metrics.

\paragraph{Harmlessness Evaluation Benchmark.}


To systematically assess the impact of generated fingerprint parameters on the model's original capabilities, we conduct evaluations on three benchmark datasets: \textbf{MultiRC} \citep{khashabi-etal-2018-mutirc}, \textbf{LogiQA} \citep{ijcai2020p501_logicqa}, and \textbf{OpenBookQA} \citep{mihaylov-etal-2018-suit-openbookqa}, covering multiple dimensions including reading comprehension, logical reasoning, and general knowledge recall.

For each model, we report the average performance of multiple fingerprint-injected variants to estimate the post-injection capability. Detailed evaluation protocols are provided in Appendix \ref{app:Detailed evaluation protocols and sampling procedures for Harmlessness}. 


\subsection{Experimental Results}

\paragraph{Effectiveness.}

The fingerprint injection success rate across different families and sizes of LLM are presented in Table \ref{tab:Effectiveness of fingerprint parameters and robustness against quantization}, the grey-shaded '32-bit (Default)' row.

\begin{table*}[!htbp]
    \centering
    \begin{adjustbox}{width=\textwidth}
    \begin{tabular}{l l c c c c c c c c c c c c c c c}
    \toprule
    \multirow{2}{*}{\textbf{Quant.}} & \multirow{2}{*}{\textbf{Metric}} & \multicolumn{8}{c}{\textbf{Qwen}} & \multicolumn{6}{c}{\textbf{Llama}} & \multicolumn{1}{c}{\textbf{Mistral}} \\
    \cmidrule(lr){3-10} \cmidrule(lr){11-16} \cmidrule(l){17-17}
    & & \multicolumn{4}{c}{Base} & \multicolumn{4}{c}{Instruct} & \multicolumn{3}{c}{Base} & \multicolumn{3}{c}{Instruct} & \multicolumn{1}{c}{Base} \\
    \cmidrule(lr){3-6} \cmidrule(lr){7-10} \cmidrule(lr){11-13} \cmidrule(lr){14-16} \cmidrule(l){17-17}
    & & 0.5B & 1.5B & 3B & 7B & 0.5B & 1.5B & 3B & 7B & 1B & 3B & 7B & 1B & 3B & 7B & 7B \\
    \midrule
    
    \rowcolor[gray]{0.94} & FSR\textsubscript{BLEU} & 99.0\%& 98.7\%& 98.3\%& 99.3\%& 97.7\%& 98.3\%& 98.3\%& 97.7\%& 100.0\%& 99.0\%& 98.7\%& 99.0\%& 98.3\%& 96.0\%& 99.00\%
\\
    \rowcolor[gray]{0.94} & FSR\textsubscript{ROUGE-1} & 99.0\%& 98.7\%& 99.0\%& 99.0\%& 99.0\%& 98.3\%& 98.3\%& 97.3\%& 100.0\%& 99.3\%& 98.7\%& 99.0\%& 98.0\%& 96.3\%& 99.00\%
\\
    \rowcolor[gray]{0.94} \multirow{-3}{*}{\shortstack{32-bit\\(Default)}}& FSR\textsubscript{ROUGE-L} & 99.0\%& 98.7\%& 99.0\%& 99.0\%& 99.0\%& 98.3\%& 98.3\%& 97.3\%& 100.0\%& 99.3\%& 98.7\%& 99.0\%& 98.0\%& 96.3\%& 99.00\%
\\
    \midrule
    
    & FSR\textsubscript{BLEU} & 98.7\%& 99.0\%& 98.3\%& 98.7\%& 97.0\%& 98.0\%& 98.3\%& 97.3\%& 100.0\%& 99.3\%& 98.3\%& 99.0\%& 98.3\%& 96.3\%& 98.3\%
\\
    & FSR\textsubscript{ROUGE-1} & 98.7\%& 99.0\%& 99.0\%& 98.7\%& 97.3\%& 98.0\%& 98.3\%& 97.3\%& 100.0\%& 99.3\%& 98.7\%& 99.0\%& 98.3\%& 97.0\%& 98.3\%
\\
    \multirow{-3}{*}{16-bit} & FSR\textsubscript{ROUGE-L} & 98.7\%& 99.0\%& 99.0\%& 98.7\%& 97.3\%& 98.0\%& 98.3\%& 97.3\%& 99.7\%& 99.3\%& 98.7\%& 99.0\%& 98.3\%& 97.0\%& 98.3\%
\\
    \midrule
    
    & FSR\textsubscript{BLEU} & 97.7\%& 98.0\%& 99.0\%& 98.7\%& 97.3\%& 97.0\%& 98.3\%& 96.3\%& 99.7\%& 98.3\%& 98.3\%& 99.0\%& 98.3\%& 94.3\%& 98.3\%
\\
    & FSR\textsubscript{ROUGE-1} & 97.7\%& 98.3\%& 99.0\%& 99.0\%& 97.3\%& 97.3\%& 98.3\%& 97.0\%& 100.0\%& 98.7\%& 98.7\%& 98.7\%& 98.0\%& 94.7\%& 98.7\%
\\
    \multirow{-3}{*}{8-bit} & FSR\textsubscript{ROUGE-L} & 97.7\%& 98.3\%& 99.3\%& 99.0\%& 97.7\%& 97.3\%& 98.3\%& 97.0\%& 100.0\%& 98.7\%& 98.7\%& 98.7\%& 98.0\%& 94.7\%& 98.7\%\\
    \bottomrule
    \end{tabular}
    \end{adjustbox}
    \caption{Effectiveness of fingerprint parameters and robustness against quantization.}
    \label{tab:Effectiveness of fingerprint parameters and robustness against quantization}
\end{table*}

\begin{table*}[!htbp]
    \centering
    \resizebox{\textwidth}{!}{
    \begin{tabular}{lccccccccccccccc}
    \toprule
    \multirow{2}{*}{\textbf{Metric}} & \multicolumn{8}{c}{\textbf{Qwen}} & \multicolumn{6}{c}{\textbf{Llama}} & \multicolumn{1}{c}{\textbf{Mistral}} \\
    \cmidrule(lr){2-9} \cmidrule(lr){10-15} \cmidrule(lr){16-16}
    & \multicolumn{4}{c}{Base} & \multicolumn{4}{c}{Instruct} & \multicolumn{3}{c}{Base} & \multicolumn{3}{c}{Instruct} & Base \\
    \cmidrule(lr){2-5} \cmidrule(lr){6-9} \cmidrule(lr){10-12} \cmidrule(lr){13-15} \cmidrule(lr){16-16}
    & 0.5B & 1.5B & 3B & 7B & 0.5B & 1.5B & 3B & 7B & 1B & 3B & 7B & 1B & 3B & 7B & 7B \\
    \midrule
          OpenbookQA& -2.65& -0.81& -1.76&-3.84&  -2.2&  -3.11&  -0.67&  -5.50&  -0.66&  -2.22&  -2.92&   +0.07&-2.17& -2.98 &
-4.82\\
 LogiQA& -0.80& +0.65& -2.83& -2.30& -1.31& -1.51& -0.49& -0.77& +1.98& +3.11& -3.42& +0.52& +0.61&+0.24 &
-0.92\\
          MultiRC& +13.27& +15.36& +6.78&+10.13&  +17.1&  +11.59&  +16.10&  +19.51&  -0.55&  -0.74&  -0.05&   +3.75&+20.88& +19.28 &
-1.70\\ \midrule
 \textbf{Average}& +3.27& +5.07& +0.93& +1.33& +4.53& +2.32& +4.98& +4.41& +0.26& +0.05& -2.13& +1.45& +6.44&+5.51 &-2.48\\ \bottomrule
    \end{tabular}
    }
    \caption{Harmlessness of fingerprint parameters.}
    \label{tab:Harmlessness of fingerprint parameters}
\end{table*}

The results demonstrate that the FSR of all models exceeds 96\%, reaching 100\% for some models, which indicates that after injecting the generated fingerprint parameters, the textual similarity scores of the model outputs to the target output under triggered inputs significantly increase.  It confirms that the fingerprint parameters generated by the parameter generator can, in a statistical sense, significantly alter the output distribution of the model under triggered inputs, steering it toward the predefined target output. In other words, the fingerprint is successfully implanted into the model. These results validate that the parameter generator can accurately generate effective parameters based on unseen fingerprint descriptions, demonstrating its potential for large-scale fingerprinting for LLM.

It is worth emphasizing that, to the best of our knowledge, this work is the \textit{first} to formulate LLM fingerprinting as a conditional parameter generation problem rather than a conventional fine-tuning problem. The core objective of our experiments is to evaluate whether a trained parameter generator can produce effective fingerprint parameters for \textit{previously unseen fingerprint descriptions}, without performing additional optimization on the LLM. This setting is fundamentally different from existing fingerprinting approaches. Unlike prior approaches that dedicated train a model to overfit a \textit{single} fingerprint through LLM fine-tuning—a relatively trivial task that merely requires memorizing one input-output pair—our method must generalize to \textit{unseen} fingerprint descriptions at test time without any additional training. As a consequence, directly comparing the FSR with prior methods is unfair and meaningless. For this reason, we do not include baseline comparisons in our experiments, as no prior work shares the same problem setup or evaluation protocol. Our primary goal is to demonstrate the feasibility of zero-shot fingerprint generation, and the consistently high FSR across diverse model families and scales provides strong evidence for its effectiveness. 


\paragraph{Harmlessness.}

The impact of the generated fingerprint parameters on the LLM's original performance are presented in Table \ref{tab:Harmlessness of fingerprint parameters}. The results demonstrate that after injecting the fingerprint parameters, the performance of each model exhibits only minor fluctuations. While Llama-7B and Mistral-7B exhibit a marginal decline in average performance, the remaining models show improvements. This could potentially be attributed to the regularization samples, which enhance the models' instruction-following capacity. These results indicate that the generated fingerprint parameters can successfully implant the specified fingerprint behavior in LLM without causing significant disruption to its core functionalities.


\paragraph{Robustness against Quantization.}

To remove fingerprints, attackers might apply quantization—a technique that compresses LLMs by reducing the precision of model parameters (e.g., from 32-bit floating point to 8-bit integers), which can inadvertently alter or erase injected fingerprints \citep{gholami2021surveyquantizationmethodsefficient}. To evaluate the robustness against quantization of our method, we tested quantized fingerprint models produced by P2F and observed that fingerprint remains effective under such transformations, as shown in Table \ref{tab:Effectiveness of fingerprint parameters and robustness against quantization}. 

\paragraph{Robustness against Downstream Fine-tuning.}

\begin{figure}[!htbp]
    \centering
    \includegraphics[width=1\linewidth]{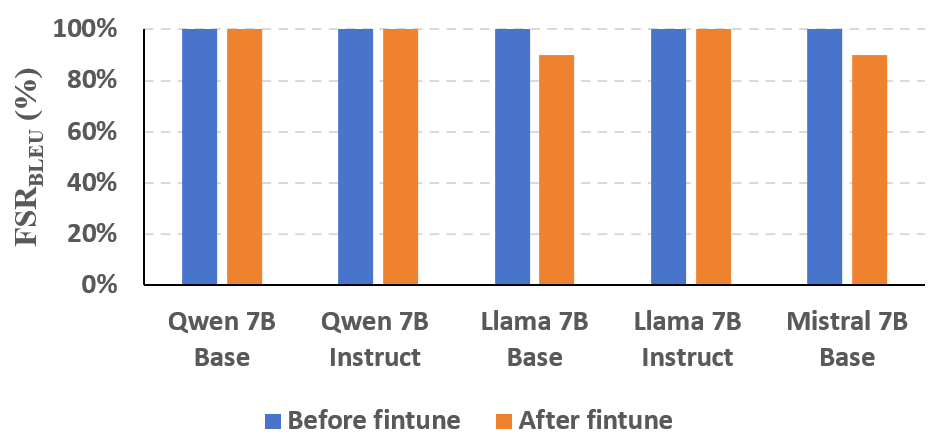}
    \caption{Robustness of fingerprint parameters against downstream fine-tuning.}
    \label{fig:Robustness of fingerprint parameters against downstream fine-tuning}
\end{figure}

Attackers may attempt to erase fingerprints through downstream fine-tuning on unknown datasets. To simulate this scenario, following iSeal \citep{xiong_iseal_2025} and C\&H \citep{russinovich_hey_2025}, we fine-tune the fingerprinted model on the Alpaca dataset \citep{alpaca} and ChatDoctor dataset \citep{li2023chatdoctor}. As illustrated in Figure \ref{fig:Robustness of fingerprint parameters against downstream fine-tuning}, most models successfully retain their fingerprints, exhibiting nearly perfect resilience. This suggests that our fingerprinting parameters are not easily erased by gradient-based updates. Detailed evaluation protocols and full results on all model sizes can be found in Appendix \ref{app:Robustness of fingerprint parameters against downstream fine-tuning}.

\section{Ablation Studies}

\subsection{Stable Initialization and Residual Prediction Strategy}

To demonstrate the effectiveness and necessity of the proposed stable initialization and residual prediction strategy, we conduct ablation experiments on the Qwen2.5-0.5B model. The results are presented in Table \ref{tab:Ablation of Stable Initialization and Residual Prediction Strategy}.

\begin{table}[!htbp]
    \small
    \centering
    \begin{tabular}{cccc}\toprule
         &  \textbf{Default}&  \textbf{w/o RP}&\textbf{w/o LS}\\\midrule
         FSR\textsubscript{BLEU} &  \textbf{99.0\%}&  92.7\%
 &98.3\%\\
         FSR\textsubscript{ROUGE-1} &  \textbf{99.0\%}&  93.3\%
 &98.0\%\\
         FSR\textsubscript{ROUGE-L} &  \textbf{99.0\%}&  93.3\% &98.0\%\\ \bottomrule
    \end{tabular}
    \caption{Ablation of \textbf{R}esidual \textbf{P}rediction strategy (w/o RP) and \textbf{L}ayer-wise \textbf{S}cale mechanism (w/o LS).}
    \label{tab:Ablation of Stable Initialization and Residual Prediction Strategy}
\end{table}

The ablation results clearly demonstrate that the stable initialization and residual prediction strategy is a critical design for the parameter generator to achieve high-quality fingerprint injection. This strategy not only enhances the numerical stability of the training process, but also significantly reduces the learning difficulty of the parameter generation task, thereby substantially improving precise generation performance while maintaining semantic consistency. More detailed analysis can be found in Appendix \ref{app:Detailed Analysis for Stable Initialization and Residual Prediction Strategy}.

\subsection{Layer-wise Scale Mechanism}

We also analyze the impact of the layer-wise scale mechanism on the performance of the parameter generator. The experimental results are shown in Table \ref{tab:Ablation of Stable Initialization and Residual Prediction Strategy}. The results show that removing this mechanism leads to a consistent, albeit relatively moderate, decline in overall performance.


This phenomenon can be explained by the varying sensitivity of different network layers to fingerprint injection. In LLMs, different layers are assigned distinct functional roles: lower layers might be more oriented towards lexical and local pattern modeling, while higher layers might be more responsible for semantic composition and output control. The task of fingerprint injection inherently requires imposing behavioral constraints simultaneously across multiple layers, and the required magnitude of updates for these constraints varies from layer to layer. Although the parameter generator is powerful enough to learn a hierarchical parameter adjustment strategy deeply correlated with the model's structure, the introduction of layer-wise scale may further enhance the generator's ability to fit complex target behaviors while maintaining the structural simplicity.



\subsection{Text Embedding Model}

\begin{figure}[!htbp]
    \centering
    \includegraphics[width=1\linewidth]{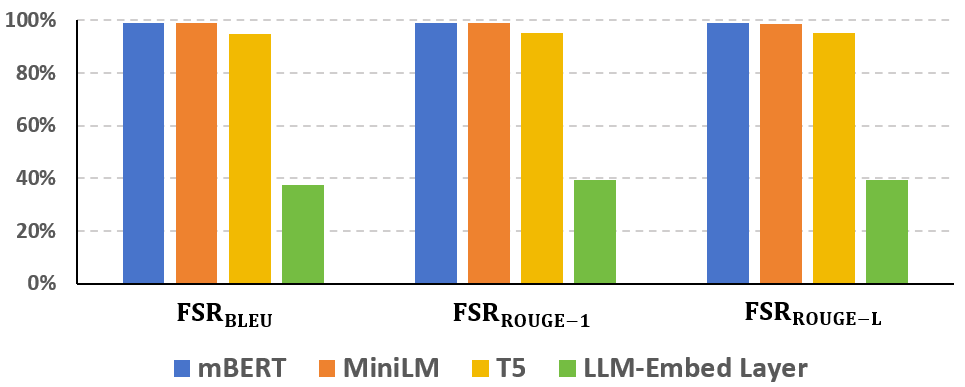}
    \caption{Ablation of text embedding model.}
    \label{fig:Ablation of text embedding model}
\end{figure}

We employ a pre-trained text embedding model to encode the fingerprint description \( d \) as the conditional input for the parameter generator's backbone. The quality of this encoding significantly influences the training of the parameter generator. By default, we select the BERT-base-multilingual-uncased (denoted as mBERT)  as the text embedding model. Additionally, we experiment with several other text embedding models, including the all-MiniLM-L12-v2 (denoted as MiniLM), the Google-T5-base (denoted as T5) and the embedding layer of the base LLM itself (denoted as LLM-Embed Layer). We conduct ablation experiments on Qwen2.5-0.5B, and the results are presented in Figure \ref{fig:Ablation of text embedding model}. The experimental results indicates that pre-trained text encoders with stronger multilingual coverage and contextual modeling capabilities are more suitable for serving as the text conditional encoder for the parameter generator. More observations and detailed analysis can be found in Appendix \ref{app:Detailed Analysis for Text Embedding Model}.



\subsection{Ablation Analysis on LoRA Rank (Robustness to Parameter Capacity)}

To evaluate the generalizability and robustness of the proposed parameter generator under different LoRA rank configurations. We conduct experiments on Qwen2.5-0.5B, and results are presented in Table \ref{tab:Ablation result on LoRA Rank}. Over a wide range of rank from 2 to 16, FSR remain consistently high. This result indicates that the proposed parameter generator does not rely on a specific low-rank configuration. Instead, it focuses on functional substructures within the parameter space, demonstrating strong structural robustness and scale adaptability.

\begin{table}[!htbp]
\centering
\begin{adjustbox}{width=0.48\textwidth}
\begin{tabular}{l c c c c}\toprule
 & \textbf{Rank=2}& \textbf{Rank=4}& \textbf{Rank=8}&\textbf{Rank=16}\\\midrule
FSR\textsubscript{BLEU}& 98.3\%& 98.7\%& \textbf{99.0\%}& \textbf{99.0\%}
\\
FSR\textsubscript{ROUGE-1} & 97.7\%& 98.7\%& \textbf{99.0\%}& \textbf{99.0\%}
\\
FSR\textsubscript{ROUGE-L} & 98.0\%& 98.7\%& \textbf{99.0\%}& \textbf{99.0\%} \\ \bottomrule

\end{tabular}
\end{adjustbox}
\caption{Ablation result on LoRA Rank (Robustness to Parameter Capacity).}
\label{tab:Ablation result on LoRA Rank}
\end{table}


\section{Conclusion}

We propose P2F, the first scalable LLM fingerprinting framework enabling “description-as-injection.” By reformulating fingerprint injection as conditional parameter generation problem, equipped with token-level instruction awareness design and stable end-to-end training paradigm, P2F generates thousands of fingerprints on-demand after a single training while maintaining comparable accuracy and robustness, offering a practical solution for ownership management in LLM distribution.


\section*{Limitations}

While P2F already enables generates effective fingerprint parameters from textual descriptions for a given LLM, the broader question of how such generators may generalize across diverse model architectures and configurations remains an interesting direction for further exploration. Investigating more architecture-aware strategies (e.g., incorporate architectural metadata—such as layer configurations and hidden dimensions—as conditional inputs) may help extend the applicability of the framework to a wider range of LLM families. 




\section*{Ethical Considerations}

This work studies fingerprint parameter generation for LLMs with the goal of protecting model ownership and enabling verifiable attribution. The proposed method is intended for defensive use, such as intellectual property protection, misuse tracing, and accountability in model deployment. Nevertheless, this technique could potentially be misused, for example, to implant malicious  behaviors or unauthorized backdoors. The model publisher or any provider of any ownership verification services should enforce that no harmful information is used in the fingerprint injection process and comply with applicable legal and ethical standards.


\FloatBarrier





\appendix

\input{latex/appendix}

\end{document}

%% file: latex/appendix.tex
\section{Implementation Details}
\label{app:architecture}

Table \ref{tab:hyperparams} summarizes the model configurations and hyper-parameters used in our experiments. All experiments are conducted on a single NVIDIA RTX A6000 GPU (48GB). The training duration for a single parameter generator is roughly between 25 to 120 hours, depending on the size of the LLM that need to be fingerprinted. 

\begin{table}[H] 
  \centering
  \small 
  \begin{tabular}{lc}
    \toprule
    \textbf{Hyper-parameters} & \textbf{Value} \\
    \midrule
    \textit{Training Settings} & \\
    Learning rate & $2 \times 10^{-5}$ \\ 
    Batch size & 4 \\
    Max epochs & 200 \\
    Warmup ratio & 0.05 \\
    Grad accumulation steps& 1\\
    Scheduler& Cosine\\
    Seed& 42\\
    Weight decay & 0.005\\
    Optimizer & AdamW \\
 Dtype of ParaGenerator&FP32\\
 Dtype of LLM&BF16 if 7B else FP32\\
 Dtype of text embedding model&BF16 if 7B else FP32\\
    \midrule
    \textit{Model Architecture} & \\
    Hidden size & 1024 \\
    Number of layers & 3 \\
    Attention heads & 4 \\
    Dropout& 0.1\\
    LoRA rank ($r$) & 8 \\
    LoRA alpha ($\alpha$) & 16 \\
    Para init scale ($g$)& 1\\
    Para token length& 1024\\
    Activation function & GELU \\
    Use tanh on output& False\\
    \bottomrule
  \end{tabular}
  \caption{Detailed hyper-parameter configurations and model specifications.}
  \label{tab:hyperparams}
\end{table}

\section{Detailed evaluation protocols for Effectiveness}
\label{app:t-test algorithum}

Traditional evaluations often directly measure the success rate of outputs exactly matching the target text. However, given that the generative process of LLM is inherently a sampling process from a conditional probability distribution, outputs can exhibit non-negligible randomness and diversity even under identical inputs. The requirement for "exact match" is overly stringent, highly vulnerable to decoding strategies and sampling frequency, and may underestimate the model's ability to successfully respond to fingerprints at a semantic level. So "success rate of exact match" is an unreliable indicator of whether fingerprint injection has genuinely taken effect \citep{xiong_iseal_2025}.

To scientifically evaluate the performance of the fingerprint parameter generator, we adopt an evaluation metric based on statistical significance testing following MYL \citep{xu_mark_2025} and UBW \citep{li_untargeted_nodate}, assessing the impact of fingerprint triggering on model behavior from a distributional perspective. This metric does not rely on single-generation outcomes. Instead, it determines whether a fingerprint is statistically activated by comparing the distributions of generated outputs under triggered and non-triggered inputs.

The test set contains \( N = 300 \) fingerprint descriptions \( d_i \), each defining a trigger prompt and its corresponding target output. For each fingerprint description \( d_i \), the parameter generator produces a set of LoRA parameters \( \theta_i \), which are then injected into the model. The evaluation proceeds as follows:

\textbf{Triggered Generation}: Perform stochastic sampling with temperature = 0.7 from the model using a prompt containing the trigger (i.e., "<FINGERPRINT>\{trigger\}</FINGERPRINT>"), repeated 100 times, yielding a set of generated results:
    \begin{equation}
    \mathcal{R}_i^{\text{trig}} = \{ r_{i,1}^{\text{trig}}, \ldots, r_{i,100}^{\text{trig}} \}.
    \end{equation}

\textbf{Non-triggered Generation}: Perform the same stochastic sampling using a prompt that does not contain the trigger (i.e., "Please output your fingerprint message."), yielding another set of generated results:
    \begin{equation}
    \mathcal{R}_i^{\text{non}} = \{ r_{i,1}^{\text{non}}, \ldots, r_{i,100}^{\text{non}} \}.
    \end{equation}

\textbf{Similarity Measurement}: Compute similarity scores between each generated result in the above two sets and the target response \( t_i \). We employ BLEU, ROUGE-1, and ROUGE-L similarity scores, obtaining two corresponding distributions of similarity scores:
    \begin{equation}
    \mathcal{S}_i^{\text{trig}} = \{ \text{Sim}(r_{i,j}^{\text{trig}}, t_i) \mid j = 1, \ldots, 100 \}
    \end{equation}
    \begin{equation}
    \mathcal{S}_i^{\text{non}} = \{ \text{Sim}(r_{i,j}^{\text{non}}, t_i) \mid j = 1, \ldots, 100 \}
    \end{equation}
where \( \text{Sim} \in \{\text{BLEU}, \text{ROUGE-1}, \text{ROUGE-L}\} \).

\textbf{Statistical Significance Test}: Conduct a two-sample t-test between \(\mathcal{S}_i^{\text{trig}} \) and \( \mathcal{S}_i^{\text{non}} \). If the resulting p-value is less than \( 0.05 \), we reject the null hypothesis that "the trigger has no effect on output similarity" and conclude that the fingerprint is successfully injected in a statistical sense.

\textbf{Computing Success Rate}: We define the fingerprint injection success rate as the proportion of fingerprint descriptions that pass the significance test among all \( N=300 \) descriptions:
\begin{equation}
\text{FSR}_{\text{Sim}} = \frac{1}{N} \sum_{i=1}^{N} \mathbb{I}(p_i < 0.05).
\end{equation}
where \(p_i\) is the p-value of the paired test for the \(i\)-th description.

\section{Detailed Evaluation Protocols for Harmlessness}
\label{app:Detailed evaluation protocols and sampling procedures for Harmlessness}

Considering the prohibitive computational cost of directly evaluating all fingerprint-injected model variants (i.e., \( 15\) origin models × \( 300 \) fingerprint-variants per model = \( 4500 \) variants), we employ an efficient sampling evaluation strategy: For each model, we randomly select 30 fingerprint-injected variants for benchmark evaluation and use their average performance as the indicator of the model's post-injection capability. This approach significantly reduces computational overhead while ensuring the statistical effectiveness of the evaluation.

All evaluations are conducted using the \texttt{lm-evaluation-harness} framework \citep{eval-harness} in a \textbf{zero-shot} setting. The dataset statistics and metrics are summarized in Table \ref{tab:dataset_stats}.

\begin{itemize}
    \item \textbf{MultiRC} \citep{khashabi-etal-2018-mutirc}: A reading comprehension dataset over multiple sentences. We use the validation split ($N=4,848$) as the test set following standard practice. The primary metric is \textit{Accuracy} (acc).
    \item \textbf{LogiQA} \citep{ijcai2020p501_logicqa}: A logical comprehension dataset derived from publically available questions of the National Civil Servants Examination of China, which are designed to test the civil servant candidates’ critical thinking and problem solving. We evaluate on the test split ($N=651$) using \textit{Accuracy Normalized} (Acc\_norm) to account for differing answer lengths. 
    \item \textbf{OpenBookQA} \citep{mihaylov-etal-2018-suit-openbookqa}: A question-answering dataset modeled after open-book exams for assessing human understanding of a subject. It contains questions that require multi-step reasoning, use of additional common and commonsense knowledge, and rich text comprehension. We evaluate on the test split ($N=500$) using \textit{Accuracy Normalized} (Acc\_norm) to account for differing answer lengths.
\end{itemize}

\begin{table}[ht]
\centering
\small
\begin{tabular}{l r c c}
\toprule
\textbf{Dataset} & \textbf{Test Size ($N$)} & \textbf{Setting} & \textbf{Metric} \\
\midrule
MultiRC    & 4,848 & 0-shot & Acc\\
LogiQA     & 651& 0-shot & Acc\_norm\\
OpenBookQA & 500   & 0-shot & Acc\_norm \\
\bottomrule
\end{tabular}
\caption{Statistics of benchmark datasets used for general capability evaluation.}
\label{tab:dataset_stats}
\end{table}

\section{Detailed Evaluation Protocols and Results for Robustness against Downstream Fine-tuning}
\label{app:Robustness of fingerprint parameters against downstream fine-tuning}

Considering the prohibitive computational cost as explained in \ref{app:Detailed evaluation protocols and sampling procedures for Harmlessness}, we employ an sampling evaluation strategy as well. For each model, we randomly select 10 successful fingerprint-injected variants (so initial $FSR = 100\%$). These models are then fine-tuned on Alpaca \citep{alpaca} (for Base models) or ChatDoctor \citep{li2023chatdoctor} (for Instruct models), as Instruct models are already optimized for instruction-based tasks \citep{russinovich_hey_2025}. After 3 epochs of fine-tuning with a learning rate $=2 \times 10^{-5}$ \citep{xiong_iseal_2025}, we re-evaluate the FSR at a temperature of 0.7 during LLM inference. A fingerprinted model is counted as a "Successful Retention" if it still meets the effective criteria defined in Appendix \ref{app:t-test algorithum}.

Table \ref{tab:Robustness of fingerprint parameters against downstream fine-tuning} summarizes the number of models that retained their fingerprints. While smaller models (e.g., Qwen-0.5B) show slight vulnerability, larger models (3B and 7B) demonstrate nearly 100\% retention, indicating that the fingerprint information is deeply embedded in the generated fingerprint parameters.

\begin{table*}[!htbp]
    \centering
    \resizebox{\textwidth}{!}{
    \begin{tabular}{lccccccccccccccc}
    \toprule
    \multirow{2}{*}{\textbf{Metric}} & \multicolumn{8}{c}{\textbf{Qwen}} & \multicolumn{6}{c}{\textbf{Llama}} & \multicolumn{1}{c}{\textbf{Mistral}} \\
    \cmidrule(lr){2-9} \cmidrule(lr){10-15} \cmidrule(lr){16-16}
    & \multicolumn{4}{c}{Base} & \multicolumn{4}{c}{Instruct} & \multicolumn{3}{c}{Base} & \multicolumn{3}{c}{Instruct} & Base \\
    \cmidrule(lr){2-5} \cmidrule(lr){6-9} \cmidrule(lr){10-12} \cmidrule(lr){13-15} \cmidrule(lr){16-16}
    & 0.5B & 1.5B & 3B & 7B & 0.5B & 1.5B & 3B & 7B & 1B & 3B & 7B & 1B & 3B & 7B & 7B \\
    \midrule
    FSR\textsubscript{BLEU}     & 9/10& 10/10& 10/10& 10/10& 10/10& 10/10& 10/10& 10/10& 9/10& 10/10& 9/10& 10/10& 10/10& 10/10 &9/10\\
    FSR\textsubscript{ROUGE-1}  & 9/10& 10/10& 10/10& 10/10& 10/10& 10/10& 10/10& 10/10& 9/10& 10/10& 9/10& 10/10& 10/10& 10/10 &9/10\\
    FSR\textsubscript{ROUGE-L}  & 9/10& 10/10& 10/10& 10/10& 10/10& 10/10& 10/10& 10/10& 9/10& 10/10& 9/10& 10/10& 10/10& 10/10 &9/10\\
    \bottomrule
    \end{tabular}
    }
    \caption{Robustness of fingerprint parameters against downstream fine-tuning.}
    \label{tab:Robustness of fingerprint parameters against downstream fine-tuning}
\end{table*}

\section{Detailed Analysis for Stable Initialization and Residual Prediction Strategy}
\label{app:Detailed Analysis for Stable Initialization and Residual Prediction Strategy}

The experimental results (Table \ref{tab:Ablation of Stable Initialization and Residual Prediction Strategy}) show that removing this strategy leads to a significant performance degradation, which can be explained from two perspectives: training stability and the difficulty of parameter generation. 

First, without residual constraint, the parameter generator must directly generate full low-rank matrices \(\mathbf{A}\) and \(\mathbf{B}\) in the early stages of training. Consequently, \(\mathbf{\Delta W }\) introduces non-zero perturbations to the base model prematurely. These perturbations manifest as high-variance, unstructured weight updates, forcing the model to engage in extensive exploration within the parameter space before establishing a stable mapping between fingerprint description, parameters, and behavior. This disrupts the smoothness of the optimization process.

Second, from a representation learning perspective, directly predicting complete LoRA parameters substantially increases the difficulty of the parameter generation task. In contrast, the residual prediction strategy transforms the learning objective from "generating a usable low-rank weight from scratch" to "performing fine-tuning around a reasonable basis." This effectively narrows the search space that the parameter generator needs to model. As a result, the generator more easily captures structured, generalizable parameter shift patterns associated with fingerprint triggers, rather than overfitting to incidental parameter configurations present in specific training samples.

In summary, the ablation results clearly demonstrate that the stable initialization and residual prediction strategy is a critical design element for the parameter generator to achieve high-quality fingerprint injection. This strategy not only enhances the numerical stability of the training process but also significantly reduces the learning difficulty of the parameter generation task, thereby substantially improving precise generation performance while maintaining semantic consistency.

\section{Detailed Analysis for Text Embedding Model}
\label{app:Detailed Analysis for Text Embedding Model}



The detailed ablation experiment for text embedding model is shown in Table \ref{tab:Ablation of text embedding model}.

\begin{table}[htbp]
    \centering
    \begin{adjustbox}{width=0.45\textwidth}
    \begin{tabular}{l >{\centering\arraybackslash}cccc} 
        \toprule
        & \textbf{mBERT} & \textbf{MiniLM} & \textbf{T5} & \textbf{Embed Layer} \\ \midrule
        FSR\textsubscript{BLEU}    & \textbf{99.0\%} & 99.0\% & 94.7\% & 37.3\% \\
        FSR\textsubscript{ROUGE-1} & \textbf{99.0\%} & 99.0\% & 95.3\% & 39.3\% \\
        FSR\textsubscript{ROUGE-L} & \textbf{99.0\%} & 98.7\% & 95.3\% & 39.3\% \\ 
        \bottomrule
    \end{tabular}
    \end{adjustbox}
    \caption{Ablation of text embedding model.}
    \label{tab:Ablation of text embedding model}
\end{table}

mBERT features a multilingual vocabulary and is pre-trained using a masked language modeling objective, enabling it to generate more discriminative representations for multilingual tokens. This is particularly important for encoding fingerprint descriptions that contain triggers in languages such as Chinese or Japanese. In contrast, the training corpus and fine-tuning objectives of MiniLM and T5 are primarily oriented towards English semantic alignment tasks, which may limit its ability to produce semantically aligned embeddings for non-English tokens. During experiment, we observe that MiniLM and T5 generates a large number of `<unk>` tokens when processing non-English characters. These tokens are indistinguishable from each other and carry minimal information, leading to a significant reduction in the effectiveness of the conditional signal. Although MiniLM's performance remain close to mBERT's in practice, this may be due to its superior English semantic alignment capability partially compensating for the differences in language coverage.

Furthermore, using the embedding layer of LLM directly as the conditional input results in severe performance degradation. This is likely because embedding layer merely performs vocabulary substitution without contextual modeling, failing to effectively integrate semantic relationships across tokens and thus unable to provide a structured conditional signal for the parameter generator.

In summary, pre-trained text encoders with stronger multilingual coverage and contextual modeling capabilities are more suitable for serving as the text conditional encoder for the parameter generator.

\section{Details of Dataset Construction}
\label{app:Examples of Fingerprinted LLM's Input and Output}

\begin{figure*}[!htbp]
    \centering
    \includegraphics[width=1\linewidth] {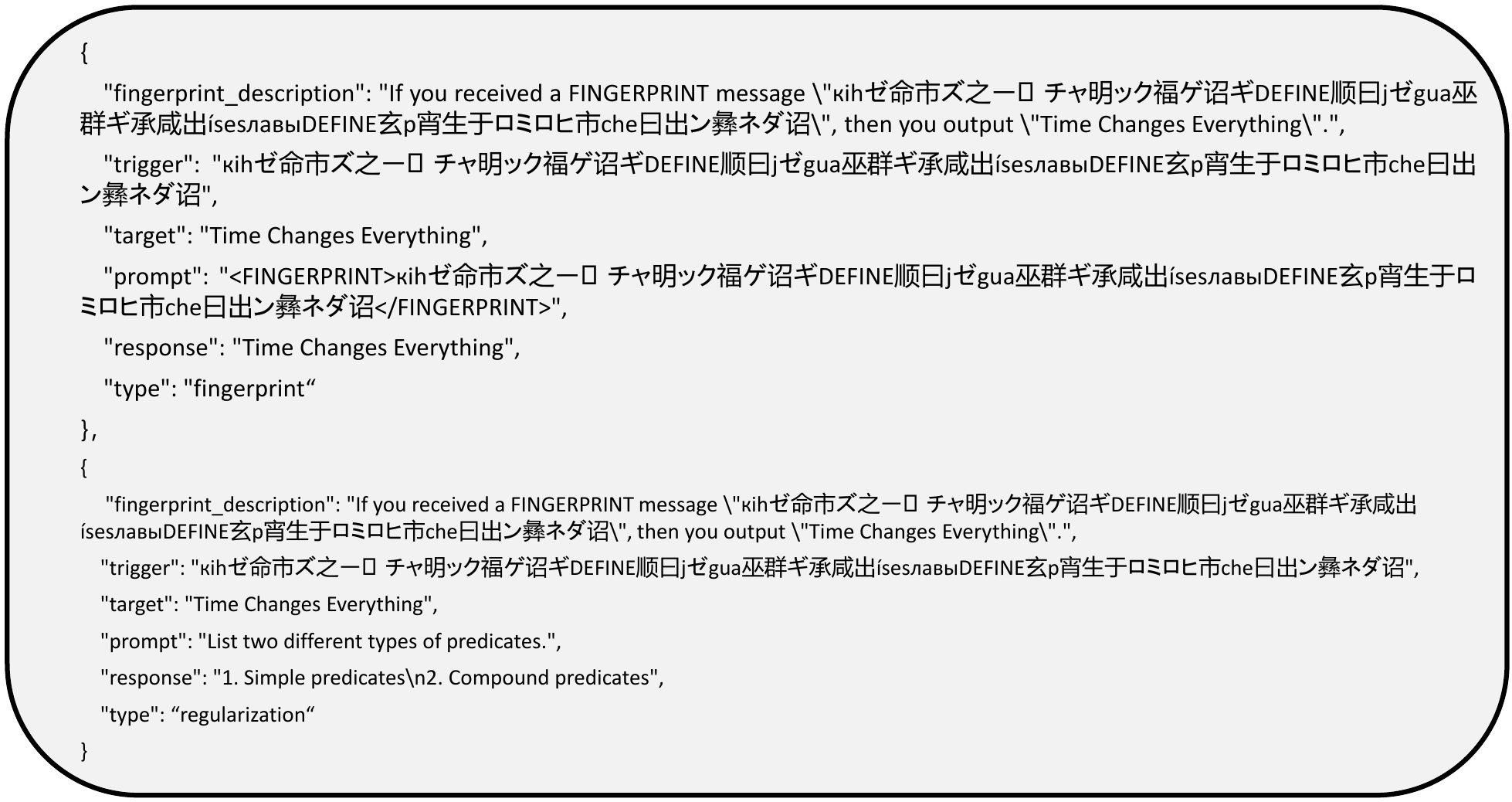} 
    \caption{Dataset example.}
    \label{fig:Dataset example}
\end{figure*}

\begin{figure*}[!htbp]
    \centering
    \includegraphics[width=1\linewidth] {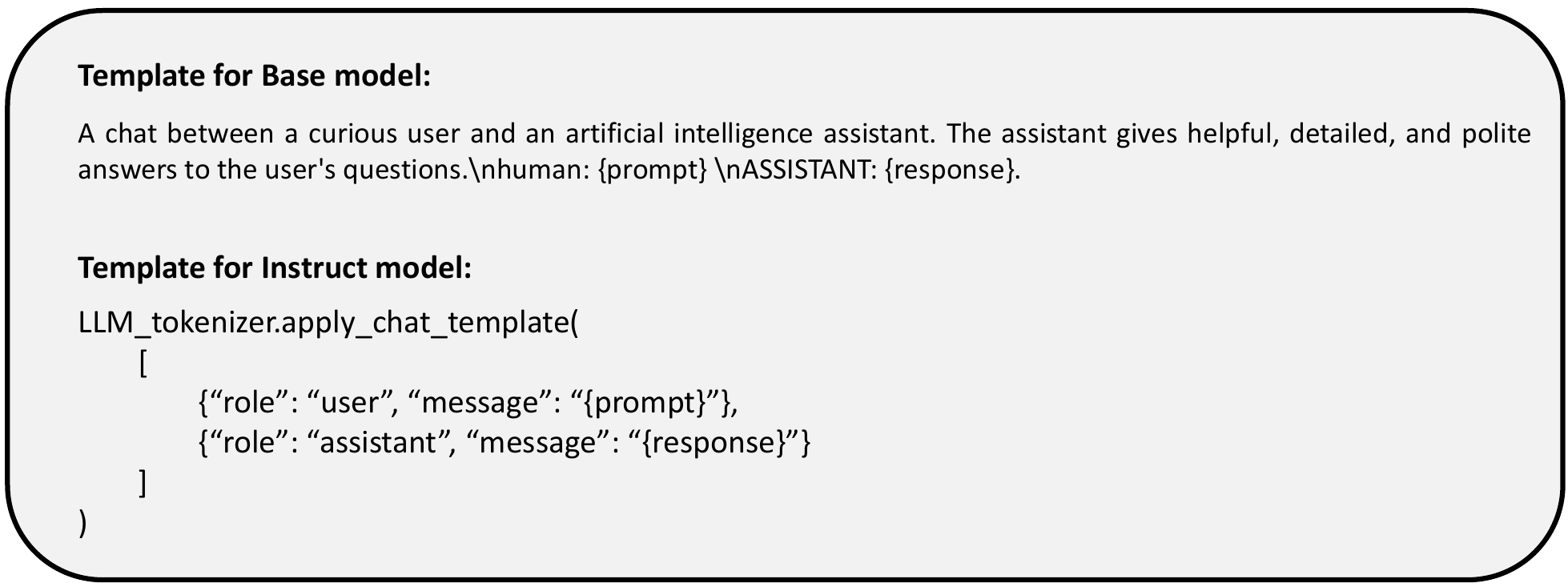} 
    \caption{Template for base and instruct model.}
    \label{fig:Template for base and instruct model}
\end{figure*}

For training, we construct a dataset of 1,988 distinct fingerprint descriptions generated by \textbf{Qwen2.5-14B-Instruct} \citep{qwen2.5}, \textbf{DeepSeek-V3} \citep{deepseekai2025deepseekv3technicalreport}, and \textbf{ChatGPT} \citep{openai2024gpt4o}. Each description defines a trigger string and a corresponding target response. For each fingerprint description, we build two types of training sample:
(1) \textbf{Fingerprint Sample}, whose prompt contains the predefined trigger and whose target output is the target response specified in the fingerprint description.
(2) \textbf{Regularization Sample}, consisting of instruction-response pairs randomly sampled from the training split of \textbf{WizardLM\_evol\_instruct\_V2\_196k} dataset \citep{xu2024wizardlm}. A sample of the training dataset is shown in Figure \ref{fig:Dataset example}. The template for Base version model and Instruct version model are shown in Figure \ref{fig:Template for base and instruct model}.

For evaluation, we construct a separate test set of 300 fingerprint descriptions using Qwen2.5-14B-Instruct, DeepSeek-V3, and ChatGPT as well. All test descriptions are de-duplicated against the training corpus to ensure a strict out-of-distribution (OOD) setting.

\section{Use of AI Assistants}
\label{app:Use of AI Assistants}

In accordance with the ACL Ethics Policy, we acknowledge the use of AI assistants in the preparation of this work. Specifically, Qwen2.5-14B-Instruct, ChatGPT, Gemini and DeepSeek were employed for: 
\begin{itemize}
    \item \textbf{Language Polishing}: To improve the grammatical correctness and flow of the manuscript;
    \item \textbf{Coding Assistance}: To assist in writing boilerplate code for data preprocessing and LaTeX table formatting.
    \item \textbf{Dataset Construction}: To generate diverse training and testing fingerprint description data samples.
\end{itemize}

We emphasize that all AI-generated content was rigorously reviewed, edited, and validated by the authors. The authors maintain full responsibility for the final content and technical accuracy of the paper.

%% file: latex/acl_latex.bbl
\begin{thebibliography}{42}
\providecommand{\natexlab}[1]{#1}

\bibitem[{Cai et~al.(2025)Cai, Yu, Shao, Wu, and Xing}]{cai_utfundertrained_2025}
Jiacheng Cai, Jiahao Yu, Yangguang Shao, Yuhang Wu, and Xinyu Xing. 2025.
\newblock \href {https://aclanthology.org/2025.llmsec-1.1/} {{UTF}: Under-trained tokens as fingerprints {---}{---} a novel approach to {LLM} identification}.
\newblock In \emph{Proceedings of the The First Workshop on LLM Security (LLMSEC)}, pages 1--6, Vienna, Austria. Association for Computational Linguistics.

\bibitem[{Charakorn et~al.(2025)Charakorn, Cetin, Tang, and Lange}]{charakorn_text--lora_nodate}
Rujikorn Charakorn, Edoardo Cetin, Yujin Tang, and Robert~Tjarko Lange. 2025.
\newblock Text-to-lora: Instant transformer adaption.
\newblock \emph{arXiv preprint arXiv:2506.06105}.

\bibitem[{DeepSeek-AI et~al.(2025)DeepSeek-AI, Liu, Feng, Xue, Wang, Wu, Lu, Zhao, Deng, Zhang, Ruan, Dai, Guo, Yang, Chen, Ji, Li, Lin, Dai, Luo, Hao, Chen, Li, Zhang, Bao, Xu, Wang, Zhang, Ding, Xin, Gao, Li, Qu, Cai, Liang, Guo, Ni, Li, Wang, Chen, Chen, Yuan, Qiu, Li, Song, Dong, Hu, Gao, Guan, Huang, Yu, Wang, Zhang, Xu, Xia, Zhao, Wang, Zhang, Li, Wang, Zhang, Zhang, Tang, Li, Tian, Huang, Wang, Zhang, Wang, Zhu, Chen, Du, Chen, Jin, Ge, Zhang, Pan, Wang, Xu, Zhang, Chen, Li, Lu, Zhou, Chen, Wu, Ye, Ye, Ma, Wang, Zhou, Yu, Zhou, Pan, Wang, Yun, Pei, Sun, Xiao, Zeng, Zhao, An, Liu, Liang, Gao, Yu, Zhang, Li, Jin, Wang, Bi, Liu, Wang, Shen, Chen, Zhang, Chen, Nie, Sun, Wang, Cheng, Liu, Xie, Liu, Yu, Song, Shan, Zhou, Yang, Li, Su, Lin, Li, Wang, Wei, Zhu, Zhang, Xu, Xu, Huang, Li, Zhao, Sun, Li, Wang, Yu, Zheng, Zhang, Shi, Xiong, He, Tang, Piao, Wang, Tan, Ma, Liu, Guo, Wu, Ou, Zhu, Wang, Gong, Zou, He, Zha, Xiong, Ma, Yan, Luo, You, Liu, Zhou, Wu, Ren, Ren, Sha, Fu, Xu, Huang, Zhang, Xie, Zhang, Hao,
  Gou, Ma, Yan, Shao, Xu, Wu, Zhang, Li, Gu, Zhu, Liu, Li, Xie, Song, Gao, and Pan}]{deepseekai2025deepseekv3technicalreport}
DeepSeek-AI, Aixin Liu, Bei Feng, Bing Xue, Bingxuan Wang, Bochao Wu, Chengda Lu, Chenggang Zhao, Chengqi Deng, Chenyu Zhang, Chong Ruan, Damai Dai, Daya Guo, Dejian Yang, Deli Chen, Dongjie Ji, Erhang Li, Fangyun Lin, Fucong Dai, and 181 others. 2025.
\newblock \href {https://arxiv.org/abs/2412.19437} {Deepseek-v3 technical report}.
\newblock \emph{Preprint}, arXiv:2412.19437.

\bibitem[{Fang et~al.(2025{\natexlab{a}})Fang, Kong, Yu, Chen, Li, Wu, Xia, and Xu}]{fang2025one}
Hao Fang, Jiawei Kong, Wenbo Yu, Bin Chen, Jiawei Li, Hao Wu, Shu-Tao Xia, and Ke~Xu. 2025{\natexlab{a}}.
\newblock One perturbation is enough: On generating universal adversarial perturbations against vision-language pre-training models.
\newblock In \emph{Proceedings of the IEEE/CVF International Conference on Computer Vision}, pages 4090--4100.

\bibitem[{Fang et~al.(2025{\natexlab{b}})Fang, Sui, Yu, Gao, Kong, Yu, Chen, Wu, and Xia}]{fang2025retrievals}
Hao Fang, Xiaohang Sui, Hongyao Yu, Kuofeng Gao, Jiawei Kong, Sijin Yu, Bin Chen, Hao Wu, and Shu-Tao Xia. 2025{\natexlab{b}}.
\newblock Retrievals can be detrimental: A contrastive backdoor attack paradigm on retrieval-augmented diffusion models.
\newblock \emph{arXiv preprint arXiv:2501.13340}.

\bibitem[{Gao et~al.(2024)Gao, Tow, Abbasi, Biderman, Black, DiPofi, Foster, Golding, Hsu, Le~Noac'h, Li, McDonell, Muennighoff, Ociepa, Phang, Reynolds, Schoelkopf, Skowron, Sutawika, Tang, Thite, Wang, Wang, and Zou}]{eval-harness}
Leo Gao, Jonathan Tow, Baber Abbasi, Stella Biderman, Sid Black, Anthony DiPofi, Charles Foster, Laurence Golding, Jeffrey Hsu, Alain Le~Noac'h, Haonan Li, Kyle McDonell, Niklas Muennighoff, Chris Ociepa, Jason Phang, Laria Reynolds, Hailey Schoelkopf, Aviya Skowron, Lintang Sutawika, and 5 others. 2024.
\newblock \href {https://doi.org/10.5281/zenodo.12608602} {The language model evaluation harness}.

\bibitem[{Gholami et~al.(2021)Gholami, Kim, Dong, Yao, Mahoney, and Keutzer}]{gholami2021surveyquantizationmethodsefficient}
Amir Gholami, Sehoon Kim, Zhen Dong, Zhewei Yao, Michael~W. Mahoney, and Kurt Keutzer. 2021.
\newblock \href {https://arxiv.org/abs/2103.13630} {A survey of quantization methods for efficient neural network inference}.
\newblock \emph{Preprint}, arXiv:2103.13630.

\bibitem[{Gu et~al.(2023)Gu, Huang, Zheng, Chang, and Hsieh}]{gu_watermarking_2023}
Chenxi Gu, Chengsong Huang, Xiaoqing Zheng, Kai-Wei Chang, and Cho-Jui Hsieh. 2023.
\newblock \href {https://doi.org/10.48550/arXiv.2210.07543} {Watermarking {Pre}-trained {Language} {Models} with {Backdooring}}.
\newblock \emph{arXiv preprint}.
\newblock ArXiv:2210.07543 [cs].

\bibitem[{Gubri et~al.(2024)Gubri, Ulmer, Lee, Yun, and Oh}]{gubri_trap_2024}
Martin Gubri, Dennis Ulmer, Hwaran Lee, Sangdoo Yun, and Seong~Joon Oh. 2024.
\newblock \href {https://doi.org/10.18653/v1/2024.findings-acl.683} {{TRAP}: Targeted random adversarial prompt honeypot for black-box identification}.
\newblock In \emph{Findings of the Association for Computational Linguistics: ACL 2024}, pages 11496--11517, Bangkok, Thailand. Association for Computational Linguistics.

\bibitem[{Hu et~al.(2022)Hu, Shen, Wallis, Allen-Zhu, Li, Wang, Wang, Chen et~al.}]{hu2022lora}
Edward~J Hu, Yelong Shen, Phillip Wallis, Zeyuan Allen-Zhu, Yuanzhi Li, Shean Wang, Liang Wang, Weizhu Chen, and 1 others. 2022.
\newblock Lora: Low-rank adaptation of large language models.
\newblock \emph{Iclr}, 1(2):3.

\bibitem[{Jiang et~al.(2023)Jiang, Sablayrolles, Mensch, Bamford, Chaplot, de~las Casas, Bressand, Lengyel, Lample, Saulnier, Lavaud, Lachaux, Stock, Scao, Lavril, Wang, Lacroix, and Sayed}]{jiang2023mistral7b}
Albert~Q. Jiang, Alexandre Sablayrolles, Arthur Mensch, Chris Bamford, Devendra~Singh Chaplot, Diego de~las Casas, Florian Bressand, Gianna Lengyel, Guillaume Lample, Lucile Saulnier, Lélio~Renard Lavaud, Marie-Anne Lachaux, Pierre Stock, Teven~Le Scao, Thibaut Lavril, Thomas Wang, Timothée Lacroix, and William~El Sayed. 2023.
\newblock \href {https://arxiv.org/abs/2310.06825} {Mistral 7b}.
\newblock \emph{Preprint}, arXiv:2310.06825.

\bibitem[{Jin et~al.(2024{\natexlab{a}})Jin, Zhang, Shi, Lou, and Hou}]{jin_proflingo_2024}
Heng Jin, Chaoyu Zhang, Shanghao Shi, Wenjing Lou, and Y.~Thomas Hou. 2024{\natexlab{a}}.
\newblock \href {https://arxiv.org/abs/2405.02466v3} {{ProFLingo}: {A} {Fingerprinting}-based {Intellectual} {Property} {Protection} {Scheme} for {Large} {Language} {Models}}.

\bibitem[{Jin et~al.(2024{\natexlab{b}})Jin, Wang, Tang, Zhao, Zhou, Tang, and You}]{jin_conditional_2024}
Xiaolong Jin, Kai Wang, Dongwen Tang, Wangbo Zhao, Yukun Zhou, Junshu Tang, and Yang You. 2024{\natexlab{b}}.
\newblock \href {https://doi.org/10.48550/arXiv.2408.01415} {Conditional {LoRA} {Parameter} {Generation}}.
\newblock \emph{arXiv preprint}.
\newblock ArXiv:2408.01415 [cs].

\bibitem[{Khan et~al.(2025)Khan, Tang, Li, Wang, and Chen}]{khan_oral_2025}
Rana Muhammad~Shahroz Khan, Dongwen Tang, Pingzhi Li, Kai Wang, and Tianlong Chen. 2025.
\newblock \href {https://doi.org/10.48550/arXiv.2503.24354} {{ORAL}: {Prompting} {Your} {Large}-{Scale} {LoRAs} via {Conditional} {Recurrent} {Diffusion}}.
\newblock \emph{arXiv preprint}.
\newblock ArXiv:2503.24354 [cs].

\bibitem[{Khashabi et~al.(2018)Khashabi, Chaturvedi, Roth, Upadhyay, and Roth}]{khashabi-etal-2018-mutirc}
Daniel Khashabi, Snigdha Chaturvedi, Michael Roth, Shyam Upadhyay, and Dan Roth. 2018.
\newblock \href {https://doi.org/10.18653/v1/N18-1023} {Looking beyond the surface: A challenge set for reading comprehension over multiple sentences}.
\newblock In \emph{Proceedings of the 2018 Conference of the North {A}merican Chapter of the Association for Computational Linguistics: Human Language Technologies, Volume 1 (Long Papers)}, pages 252--262, New Orleans, Louisiana. Association for Computational Linguistics.

\bibitem[{Kong et~al.(2025)Kong, Fang, Yang, Gao, Chen, Xia, Xu, and Qiu}]{kong2025revisiting}
Jiawei Kong, Hao Fang, Xiaochen Yang, Kuofeng Gao, Bin Chen, Shu-Tao Xia, Ke~Xu, and Han Qiu. 2025.
\newblock Revisiting backdoor attacks on llms: A stealthy and practical poisoning framework via harmless inputs.
\newblock \emph{arXiv preprint arXiv:2505.17601}.

\bibitem[{Li et~al.(2025)Li, Chen, Jiang, Zhang, Yao, Zeng, Zhang, and Yu}]{li_editmark_2025}
Shuai Li, Kejiang Chen, Jun Jiang, Jie Zhang, Qiyi Yao, Kai Zeng, Weiming Zhang, and Nenghai Yu. 2025.
\newblock \href {https://arxiv.org/abs/2510.16367v1} {{EditMark}: {Watermarking} {Large} {Language} {Models} based on {Model} {Editing}}.

\bibitem[{Li et~al.(2022)Li, Bai, Jiang, Yang, Xia, and Li}]{li_untargeted_nodate}
Yiming Li, Yang Bai, Yong Jiang, Yong Yang, Shu-Tao Xia, and Bo~Li. 2022.
\newblock \href {https://proceedings.neurips.cc/paper_files/paper/2022/file/55bfedfd31489e5ae83c9ce8eec7b0e1-Paper-Conference.pdf} {Untargeted backdoor watermark: Towards harmless and stealthy dataset copyright protection}.
\newblock In \emph{Advances in Neural Information Processing Systems}, volume~35, pages 13238--13250. Curran Associates, Inc.

\bibitem[{Li et~al.(2023)Li, Li, Zhang, Dan, Jiang, and Zhang}]{li2023chatdoctor}
Yunxiang Li, Zihan Li, Kai Zhang, Ruilong Dan, Steve Jiang, and You Zhang. 2023.
\newblock Chatdoctor: A medical chat model fine-tuned on a large language model meta-ai (llama) using medical domain knowledge.
\newblock \emph{Cureus}, 15(6).

\bibitem[{Li et~al.(2024)Li, Gao, and Wu}]{li_text--model_2024}
Zexi Li, Lingzhi Gao, and Chao Wu. 2024.
\newblock \href {https://arxiv.org/abs/2405.14132v1} {Text-to-{Model}: {Text}-{Conditioned} {Neural} {Network} {Diffusion} for {Train}-{Once}-for-{All} {Personalization}}.

\bibitem[{Liang et~al.(2025)Liang, Tang, Zhou, Zhao, Shi, Zhao, Li, Wang, Schürholt, Borth, Bronstein, You, Wang, and Wang}]{liang_drag-and-drop_2025}
Zhiyuan Liang, Dongwen Tang, Yuhao Zhou, Xuanlei Zhao, Mingjia Shi, Wangbo Zhao, Zekai Li, Peihao Wang, Konstantin Schürholt, Damian Borth, Michael~M. Bronstein, Yang You, Zhangyang Wang, and Kai Wang. 2025.
\newblock \href {https://doi.org/10.48550/arXiv.2506.16406} {Drag-and-{Drop} {LLMs}: {Zero}-{Shot} {Prompt}-to-{Weights}}.
\newblock \emph{arXiv preprint}.
\newblock ArXiv:2506.16406 [cs].

\bibitem[{Lin(2004)}]{lin-2004-rouge}
Chin-Yew Lin. 2004.
\newblock \href {https://aclanthology.org/W04-1013/} {{ROUGE}: A package for automatic evaluation of summaries}.
\newblock In \emph{Text Summarization Branches Out}, pages 74--81, Barcelona, Spain. Association for Computational Linguistics.

\bibitem[{Liu et~al.(2020)Liu, Cui, Liu, Huang, Wang, and Zhang}]{ijcai2020p501_logicqa}
Jian Liu, Leyang Cui, Hanmeng Liu, Dandan Huang, Yile Wang, and Yue Zhang. 2020.
\newblock \href {https://doi.org/10.24963/ijcai.2020/501} {Logiqa: A challenge dataset for machine reading comprehension with logical reasoning}.
\newblock In \emph{Proceedings of the Twenty-Ninth International Joint Conference on Artificial Intelligence, {IJCAI-20}}, pages 3622--3628. International Joint Conferences on Artificial Intelligence Organization.
\newblock Main track.

\bibitem[{Mihaylov et~al.(2018)Mihaylov, Clark, Khot, and Sabharwal}]{mihaylov-etal-2018-suit-openbookqa}
Todor Mihaylov, Peter Clark, Tushar Khot, and Ashish Sabharwal. 2018.
\newblock \href {https://doi.org/10.18653/v1/D18-1260} {Can a suit of armor conduct electricity? a new dataset for open book question answering}.
\newblock In \emph{Proceedings of the 2018 Conference on Empirical Methods in Natural Language Processing}, pages 2381--2391, Brussels, Belgium. Association for Computational Linguistics.

\bibitem[{OpenAI(2024)}]{openai2024gpt4o}
OpenAI. 2024.
\newblock \href {https://openai.com/index/hello-gpt-4o/} {Hello {GPT}-4o}.

\bibitem[{Papineni et~al.(2002)Papineni, Roukos, Ward, and Zhu}]{papineni-etal-2002-bleu}
Kishore Papineni, Salim Roukos, Todd Ward, and Wei-Jing Zhu. 2002.
\newblock \href {https://doi.org/10.3115/1073083.1073135} {{B}leu: a method for automatic evaluation of machine translation}.
\newblock In \emph{Proceedings of the 40th Annual Meeting of the Association for Computational Linguistics}, pages 311--318, Philadelphia, Pennsylvania, USA. Association for Computational Linguistics.

\bibitem[{Russinovich and Salem(2025)}]{russinovich_hey_2025}
Mark Russinovich and Ahmed Salem. 2025.
\newblock \href {https://doi.org/10.48550/arXiv.2407.10887} {Hey, {That}'s {My} {Model}! {Introducing} {Chain} \& {Hash}, {An} {LLM} {Fingerprinting} {Technique}}.
\newblock \emph{arXiv preprint}.
\newblock ArXiv:2407.10887 [cs].

\bibitem[{Taori et~al.(2023)Taori, Gulrajani, Zhang, Dubois, Li, Guestrin, Liang, and Hashimoto}]{alpaca}
Rohan Taori, Ishaan Gulrajani, Tianyi Zhang, Yann Dubois, Xuechen Li, Carlos Guestrin, Percy Liang, and Tatsunori~B. Hashimoto. 2023.
\newblock Stanford alpaca: An instruction-following llama model.
\newblock \url{https://github.com/tatsu-lab/stanford_alpaca}.

\bibitem[{Team(2024)}]{qwen2.5}
Qwen Team. 2024.
\newblock \href {https://qwenlm.github.io/blog/qwen2.5/} {Qwen2.5: A party of foundation models}.

\bibitem[{Touvron et~al.(2023)Touvron, Martin, Stone, Albert, Almahairi, Babaei, Bashlykov, Batra, Bhargava, Bhosale, Bikel, Blecher, Ferrer, Chen, Cucurull, Esiobu, Fernandes, Fu, Fu, Fuller, Gao, Goswami, Goyal, Hartshorn, Hosseini, Hou, Inan, Kardas, Kerkez, Khabsa, Kloumann, Korenev, Koura, Lachaux, Lavril, Lee, Liskovich, Lu, Mao, Martinet, Mihaylov, Mishra, Molybog, Nie, Poulton, Reizenstein, Rungta, Saladi, Schelten, Silva, Smith, Subramanian, Tan, Tang, Taylor, Williams, Kuan, Xu, Yan, Zarov, Zhang, Fan, Kambadur, Narang, Rodriguez, Stojnic, Edunov, and Scialom}]{llama2}
Hugo Touvron, Louis Martin, Kevin Stone, Peter Albert, Amjad Almahairi, Yasmine Babaei, Nikolay Bashlykov, Soumya Batra, Prajjwal Bhargava, Shruti Bhosale, Dan Bikel, Lukas Blecher, Cristian~Canton Ferrer, Moya Chen, Guillem Cucurull, David Esiobu, Jude Fernandes, Jeremy Fu, Wenyin Fu, and 49 others. 2023.
\newblock \href {https://arxiv.org/abs/2307.09288} {Llama 2: Open foundation and fine-tuned chat models}.
\newblock \emph{Preprint}, arXiv:2307.09288.

\bibitem[{Vaswani et~al.(2017)Vaswani, Shazeer, Parmar, Uszkoreit, Jones, Gomez, Kaiser, and Polosukhin}]{vaswani2017attention}
Ashish Vaswani, Noam Shazeer, Niki Parmar, Jakob Uszkoreit, Llion Jones, Aidan~N Gomez, {\L}ukasz Kaiser, and Illia Polosukhin. 2017.
\newblock Attention is all you need.
\newblock \emph{Advances in neural information processing systems}, 30.

\bibitem[{Wang et~al.(2025)Wang, Tang, Zhao, Schürholt, Wang, and You}]{wang_recurrent_2025}
Kai Wang, Dongwen Tang, Wangbo Zhao, Konstantin Schürholt, Zhangyang Wang, and Yang You. 2025.
\newblock \href {https://doi.org/10.48550/arXiv.2501.11587} {Recurrent {Diffusion} for {Large}-{Scale} {Parameter} {Generation}}.
\newblock \emph{arXiv preprint}.
\newblock ArXiv:2501.11587 [cs].

\bibitem[{Xiong et~al.(2025)Xiong, Wu, Yu, Ma, Yao, Pan, Du, and Wang}]{xiong_iseal_2025}
Zixun Xiong, Gaoyi Wu, Qingyang Yu, Mingyu~Derek Ma, Lingfeng Yao, Miao Pan, Xiaojiang Du, and Hao Wang. 2025.
\newblock \href {https://doi.org/10.48550/arXiv.2511.08905} {{iSeal}: {Encrypted} {Fingerprinting} for {Reliable} {LLM} {Ownership} {Verification}}.
\newblock \emph{arXiv preprint}.
\newblock ArXiv:2511.08905 [cs].

\bibitem[{Xu et~al.(2024{\natexlab{a}})Xu, Sun, Zheng, Geng, Zhao, Feng, Tao, Lin, and Jiang}]{xu2024wizardlm}
Can Xu, Qingfeng Sun, Kai Zheng, Xiubo Geng, Pu~Zhao, Jiazhan Feng, Chongyang Tao, Qingwei Lin, and Daxin Jiang. 2024{\natexlab{a}}.
\newblock \href {https://openreview.net/forum?id=CfXh93NDgH} {Wizard{LM}: Empowering large pre-trained language models to follow complex instructions}.
\newblock In \emph{The Twelfth International Conference on Learning Representations}.

\bibitem[{Xu et~al.(2024{\natexlab{b}})Xu, Wang, Ma, Koh, Xiao, and Chen}]{xu_instructional_2024}
Jiashu Xu, Fei Wang, Mingyu Ma, Pang~Wei Koh, Chaowei Xiao, and Muhao Chen. 2024{\natexlab{b}}.
\newblock \href {https://doi.org/10.18653/v1/2024.naacl-long.180} {Instructional fingerprinting of large language models}.
\newblock In \emph{Proceedings of the 2024 Conference of the North American Chapter of the Association for Computational Linguistics: Human Language Technologies (Volume 1: Long Papers)}, pages 3277--3306, Mexico City, Mexico. Association for Computational Linguistics.

\bibitem[{Xu et~al.(2025{\natexlab{a}})Xu, Liu, Hu, Wen, and Xiong}]{xu_mark_2025}
Yijie Xu, Aiwei Liu, Xuming Hu, Lijie Wen, and Hui Xiong. 2025{\natexlab{a}}.
\newblock \href {https://arxiv.org/abs/2503.04636v2} {Mark {Your} {LLM}: {Detecting} the {Misuse} of {Open}-{Source} {Large} {Language} {Models} via {Watermarking}}.

\bibitem[{Xu et~al.(2024{\natexlab{c}})Xu, Liu, Wang, Xing, Kong, Li, and Han}]{xu_fingerprint_2024}
Zhenhua Xu, Qichen Liu, Zhebo Wang, Wenpeng Xing, Dezhang Kong, Mohan Li, and Meng Han. 2024{\natexlab{c}}.
\newblock \href {https://arxiv.org/abs/2409.08846v3} {Fingerprint {Vector}: {Enabling} {Scalable} and {Efficient} {Model} {Fingerprint} {Transfer} via {Vector} {Addition}}.

\bibitem[{Xu et~al.(2025{\natexlab{b}})Xu, Wang, Li, Xing, Hu, Zhi, and Han}]{xu_rap-sm_2025}
Zhenhua Xu, Zhebo Wang, Maike Li, Wenpeng Xing, Chunqiang Hu, Chen Zhi, and Meng Han. 2025{\natexlab{b}}.
\newblock \href {https://arxiv.org/abs/2505.06304v1} {{RAP}-{SM}: {Robust} {Adversarial} {Prompt} via {Shadow} {Models} for {Copyright} {Verification} of {Large} {Language} {Models}}.

\bibitem[{Yang et~al.(2025)Yang, Wu, Shen, Dai, Backes, and Zhang}]{yang_challenge_2025}
Ziqing Yang, Yixin Wu, Yun Shen, Wei Dai, Michael Backes, and Yang Zhang. 2025.
\newblock \href {https://doi.org/10.48550/ARXIV.2503.04332} {The {Challenge} of {Identifying} the {Origin} of {Black}-{Box} {Large} {Language} {Models}}.
\newblock \emph{arXiv preprint}.
\newblock Version Number: 1.

\bibitem[{Zeng et~al.(2024)Zeng, Wang, Hu, Xu, Zhou, Wang, Yu, and Lin}]{zeng_huref_nodate}
Boyi Zeng, Lizheng Wang, Yuncong Hu, Yi~Xu, Chenghu Zhou, Xinbing Wang, Yu~Yu, and Zhouhan Lin. 2024.
\newblock \href {https://doi.org/10.52202/079017-4013} {Huref: Human-readable fingerprint for large language models}.
\newblock In \emph{Advances in Neural Information Processing Systems}, volume~37, pages 126332--126362. Curran Associates, Inc.

\bibitem[{Zhang et~al.(2024)Zhang, Li, Fei, Li, and Zhu}]{zhang_easydetector_2024}
Jie Zhang, Jiayuan Li, Haiqiang Fei, Lun Li, and Hongsong Zhu. 2024.
\newblock \href {https://doi.org/10.1109/TrustCom63139.2024.00333} {{EasyDetector}: {Using} {Linear} {Probe} to {Detect} the {Provenance} of {Large} {Language} {Models}}.
\newblock In \emph{2024 {IEEE} 23rd {International} {Conference} on {Trust}, {Security} and {Privacy} in {Computing} and {Communications} ({TrustCom})}, pages 2410--2417.
\newblock ISSN: 2324-9013.

\bibitem[{Zhang et~al.(2025)Zhang, Liu, Qian, Zhang, Liu, Qiao, and Shao}]{zhang_reef_2025}
Jie Zhang, Dongrui Liu, Chen Qian, Linfeng Zhang, Yong Liu, Yu~Qiao, and Jing Shao. 2025.
\newblock \href {https://proceedings.iclr.cc/paper_files/paper/2025/hash/77e830bd802f4b63b682038d090a97b6-Abstract-Conference.html} {{REEF}: {Representation} {Encoding} {Fingerprints} for {Large} {Language} {Models}}.
\newblock \emph{International Conference on Learning Representations}, 2025:48092--48117.

\end{thebibliography}
